\documentclass[11pt]{article}
\usepackage{graphicx}
\usepackage{mathrsfs}
\usepackage{xypic}

\usepackage{color}
\usepackage{amsfonts}
\usepackage{enumitem}
\usepackage{xypic}
\usepackage{xy}
\usepackage{amsmath}
\usepackage{booktabs}
\usepackage{threeparttable}
\usepackage{amssymb}
\usepackage{bm}
\usepackage{CJK}
\usepackage{cite}
\usepackage{romanbar}
\newtheorem{theorem}{Theorem}[section]
\newtheorem{lemma}[theorem]{Lemma}
\newtheorem{definition}[theorem]{Definition}
\newtheorem{example}[theorem]{Example}

\newtheorem{proposition}[theorem]{Proposition}

\newtheorem{corollary}[theorem]{Corollary}
\numberwithin{equation}{section}

\newcommand{\ba}{\begin{array}}
	\newcommand{\ea}{\end{array}}
\newcommand{\bt}{\begin{tabular}}
	\newcommand{\et}{\end{tabular}}
\newcommand{\btb}{\begin{table}}
	\newcommand{\etb}{\end{table}}
\newcommand{\bc}{\begin{center}}
	\newcommand{\ec}{\end{center}}
\newcommand{\bea}{\begin{eqnarray}}
	\newcommand{\eea}{\end{eqnarray}}
\newcommand{\Bea}{\begin{eqnarray*}}
	\newcommand{\Eea}{\end{eqnarray*}}
\newcommand{\beq}{\begin{equation}}
	\newcommand{\eeq}{\end{equation}}
\newcommand{\Tr}{\operatorname{Tr}}
\newcommand{\im}{\operatorname{Im}}
\newcommand{\wt}{\operatorname{wt}}
\newcommand{\PG}{\operatorname{PG}}

\newcommand{\Cay}{\operatorname{Cay}}
\newcommand{\NL}{\mathcal{NL}}

\def \bfm#1{\mbox{\boldmath$#1$}}

  \def \o {{\bfm o}}

\def \F {{\mathbb F}}

\begin{document}
	\baselineskip 16.5pt

	\title{Balanced Boolean Functions from Sparse Niho-Type Permutation Polynomials:
Few-Valued Walsh Spectra and Applications
	}
	\author{Chenxu Wang\footnote{Chenxu Wang is with the School of Mathematical Sciences, Capital Normal University, Beijing, China, 100048. Email: 2250501038@cnu.edu.cn},
    \and Jun Zhang\footnote{Jun Zhang is with the School of Mathematical Sciences, Capital Normal University, Beijing, China, 100048. Email: junz@cnu.edu.cn.}
}

\date{}
\maketitle
	\date{}
	\maketitle
	
\begin{abstract}
We study balanced Boolean functions parameterized by two-to-one mappings of the forms $P(x+x^2)$ and $P(x+x^{-1})$, where $P$ is a sparse permutation polynomial with generalized Niho-type exponents. First, for $q=2^m$ with $m$ even, we determine the complete four-valued Walsh distribution of the balanced Boolean function parameterized by $Q(x+x^2)$, where $Q(x)=x(1+x^{q-1}+x^{2(q-1)})$. We then generalize a permutation polynomial studied by Gupta and Sharma to the parameterized family $P_\lambda(x)=x^{3q-1}+\lambda x^{2q}+\lambda x^2$ over $\F_{q^2}$. We prove that $P_\lambda$ is a permutation polynomial if and only if either $\lambda=0$, or $\lambda\ne0$ and $X^3+X+\lambda^{-1}$ has no root in $\F_q$. Moreover, the number of admissible parameters is $(q+2)/3$ when $m$ is even and $(q+4)/3$ when $m$ is odd. Thus, admissible parameters exist for every $m\geq2$, and approximately one-third of the elements of $\F_q$ yield permutation polynomials. For every admissible parameter $\lambda$, we determine the complete Walsh distribution of the balanced Boolean function $f_\lambda$ parameterized by $P_\lambda(x+x^2)$ and prove that this distribution is independent of $\lambda$. For $m\geq3$, these functions have the five-valued Walsh spectrum ${q,0,-q,-2q,-3q}$, while for $m=2$ the spectrum degenerates to four values. As applications, their supports yield projective minimal binary five-weight codes,  sets with five hyperplane intersection numbers, and regular cubelike Cayley graphs with six distinct eigenvalues. Finally, for mappings of the form $P(x+x^{-1})$, we construct an infinite family of balanced Boolean functions with simple explicit trace representations.
\end{abstract}

	
	%\noindent\hrulefill \vskip 8mm	

\section{Introduction}

Let $\mathbb F_{2^n}$ denote the finite field with $2^n$ elements. A Boolean function over $\mathbb F_{2^n}$ is a mapping from $\mathbb F_{2^n}$ to $\mathbb F_2$. Boolean functions have broad applications in symmetric cryptography \cite{carlet1998codes,nyberg1993differentially}, coding theory \cite{carlet1998codes,li2021binary,li2022class,luo2008cyclic,mesnager2023several}, and sequence design \cite{canteaut2000binary,dobbertin2006niho,helleseth2005new,li2019survey,luo2016binary}. For more details, we refer the reader to \cite{carlet2021boolean,carlet1998codes,li2019survey,qu2025parametric,zhang2026balanced}.

In cryptographic applications, Boolean functions are expected to satisfy several desirable properties, including balance, high nonlinearity, high algebraic degree, and large algebraic immunity. Balance avoids statistical bias between the output values. High nonlinearity provides resistance against affine approximation and fast correlation attacks \cite{ding1991stability,meier1988fast}. High algebraic degree contributes to resistance against the Berlekamp--Massey and R{\o}njom--Helleseth attacks \cite{massey1969shift,rueppel1987products,ronjom2007new}, while large algebraic immunity is required to withstand algebraic attacks \cite{courtois2003algebraic}.

Recently, Carlet\cite{carlet2021parameterization} introduced a general parameterization method to construct Boolean functions, where the support of a Boolean function is described as the image set of a suitable vectorial function. Motivated by this idea,  Qu, Zhang, and Li\cite{qu2025parametric} proposed a parametric construction of balanced Boolean functions based on two-to-one mapping. Specifically, let $F$ be a two-to-one mapping on $\F_{2^n}$ and define the Boolean function $f_F:\F_{2^n}\rightarrow\F_2$ by
$$
f_F(x)=
\begin{cases}
1, & x\in\im(F),\\
0, & x\notin\operatorname{Im}(F).
\end{cases}
$$
Since every element of $\im(F)$ has exactly two preimages under $F$, we have $\#\im(F)=2^{n-1}$. Consequently, $f_F$ is balanced. The mapping $F$ is called a parameterization of $f_F$, and $f_F$ is said to be parameterized by $F$.

Let $P(x)$ be a permutation polynomial over $\F_{2^n}$. It is straightforward to verify that both $P(x+x^2)$ and $P(x+x^{-1})$ are two-to-one mappings over $\F_{2^n}$. Qu, Zhang, and Li \cite{qu2025parametric} investigated the cryptographic properties of the balanced Boolean functions parameterized by these mappings and established relationships between the underlying permutation polynomial $P$ and the Walsh transform, algebraic degree, and algebraic immunity of the resulting functions. They also derived explicit trace representations for several resulting functions and constructed two infinite families of balanced semi-bent functions. Subsequently, Zhang, Li, and Qu \cite{zhang2026balanced} constructed a class of balanced Boolean functions with four-valued Walsh spectra and seven classes of balanced plateaued functions, including four infinite families of semi-bent functions and a class of near-bent functions.

Motivated by the preceding work, we further investigate balanced
Boolean functions parameterized by mappings of the forms
$P(x+x^2)$ and $P(x+x^{-1})$, where $P$ is a sparse permutation
polynomial over $\F_{2^{2m}}$ with generalized Niho-type exponents.
Our objective is twofold: to construct balanced Boolean functions
with explicitly determined few-valued Walsh distributions from
$P(x+x^2)$, and to derive simple explicit trace representations for
those arising from $P(x+x^{-1})$. 

For the mappings of the form $P(x+x^2)$, we first consider the
permutation polynomial
$Q(x)=x\left(1+x^{q-1}+x^{2(q-1)}\right),
$  where $q=2^m$ with $m$  even. We determine the complete four-valued Walsh
spectrum of the balanced Boolean function $f_Q$ parameterized by $Q(x+x^2)$. In particular, its Walsh values take only the four values $q, 0, -q $ and $-2q,$ and its nonlinearity is
$\NL(f_Q)=\frac{q^2}{2}-q,$
which is equal to the nonlinearity of a semi-bent function in
$2m$ variables.

Our second contribution to the five-valued case is the construction
and analysis of a parameterized family of permutation polynomials.
Gupta and Sharma~\cite{gupta2016some} studied the permutation
property of
\[
x^2\left(1+x^{2(q-1)}+x^{3(q-1)}\right)
\]
over $\F_{q^2}$. Motivated by their work, we extend this polynomial
to the family
\[
P_\lambda(x)
=
x^{3q-1}+\lambda x^{2q}+\lambda x^2,
\]
where $ \lambda\in\F_q.$ We prove that $P_\lambda$ permutes $\F_{q^2}$ if and only if either
$\lambda=0$, or $\lambda\ne0$ and the cubic polynomial
$x^3+x+\lambda^{-1}$
has no root in $\F_q$. Moreover, we determine the exact number of
admissible parameters. More precisely, if $\Lambda_m$ denotes the
set of all $\lambda$ for which $P_\lambda$ is a permutation, using with Bluher's counting
result~\cite{bluher2004on}, we obtain
\[
|\Lambda_m|=
\begin{cases}
\dfrac{q+2}{3},&m\ \text{even},\\[2mm]
\dfrac{q+4}{3},&m\ \text{odd}.
\end{cases}
\]
Thus, admissible parameters exist for every $m\geq2$, and in fact
approximately one-third of the elements of $\F_q$ give permutation
polynomials. Hence this construction produces a substantial family
of permutation polynomials rather than isolated examples.

For every $\lambda\in\Lambda_m$, let $f_\lambda$ be the balanced
Boolean function parameterized by $P_\lambda(x+x^2)$. We determine
the complete Walsh spectrum of $f_\lambda$ and show that it is
independent of the choice of the admissible parameter $\lambda$.
For $m\geq3$, the Walsh values of all these function only takes $q, 0, -q, -2q$ and $-3q,$
and have nonlinearity
$\NL(f_\lambda)=\frac{q^2}{2}-\frac{3q}{2}.$
When $m=2$, the value $-3q$ has multiplicity zero, and then it 
degenerates to a four-valued one.

Our spectral analysis follows the general framework established by Qu, Zhang, and Li \cite{qu2025parametric}. Their result expresses the Walsh transform of a balanced Boolean function parameterized by $P(x+x^2)$ as an exponential sum associated with the underlying permutation polynomial $P$. Taking this reduction as our starting point, we exploit the generalized Niho form of $Q$ and $P_\lambda$ and apply the polar decomposition of $\mathbb F_{2^{2m}}^*$ to transform the resulting exponential sums into root-counting problems for certain low-degree equations over the unit circle $U_m$. The main technical contribution of our work lies in deriving and analyzing these equations for the specific polynomials $Q$ and $P_\lambda$. In particular, the analysis of $P_\lambda$ requires a detailed study of a family of quartic equations. By combining the resulting root distributions with the moment identities for Walsh spectra, we determine the complete four-valued and five-valued Walsh distributions of the corresponding balanced Boolean functions.

The complete five-valued Walsh distribution also gives rise to
several applications in coding theory, finite geometry, and
algebraic graph theory. From the support of $f_\lambda$, we construct
a projective minimal binary five-weight linear code. The same
support can be regarded as a point set in
$\operatorname{PG}(2m-1,2)$ having exactly five hyperplane
intersection numbers.
set. We further obtain a regular cubelike Cayley graph with exactly
six distinct eigenvalues.

Finally, using the permutation polynomials $G_k(x)=x\left(1+x^{k(2^m-1)}+x^{-k(2^m-1)}\right)$ over $\mathbb F_{2^{2m}}$, where $m$ is even, we construct an infinite family of balanced Boolean functions parameterized by $G_k(x+x^{-1})$. In contrast to the preceding spectral constructions, we directly characterize the image sets of these mappings and derive simple explicit trace representations for all the resulting functions.

The remainder of this paper is organized as follows. Section~\ref{sec2}
recalls the necessary preliminaries on Boolean functions, Walsh
transforms, two-to-one mappings, generalized Niho-type exponents,
and the unit circle. Section~\ref{sec3} studies balanced Boolean functions
parameterized by $P(x+x^2)$. It first presents a four-valued
construction and then develops the parameterized family
$P_\lambda$, determines its admissible parameters, and establishes
the complete five-valued Walsh distribution of the associated
Boolean functions. Section~\ref{sec5} presents
applications to projective minimal codes, finite geometry,
and cubelike Cayley graphs. Section~\ref{sec4} considers balanced Boolean functions
parameterized by $P(x+x^{-1})$ and derives explicit trace
representations for the resulting functions.  Finally, Section~\ref{sec6} concludes the paper
and discusses possible directions for further research.

\section{Preliminaries}\label{sec2}
Throughout this paper, unless otherwise stated, we use the following notation.
\begin{itemize}[label=$\bullet$, leftmargin=1.8em, itemsep=0pt, topsep=0pt, parsep=0pt, partopsep=0pt]
\item $\F_{2^n}$: the finite field with $2^n$ elements; 
\item $\F_{2^n}^*$: the multiplicative group $\F_{2^n}\setminus\{0\}$; \item for a finite set $E$, $\#E$ denotes its cardinality; 
\item for a set $E$ and a subset $S\subseteq E$, $E\setminus S$ denotes the set of elements in $E$ but not in $S$; 
\item $\bar{x}$: the $2^m$-th power $x^{2^m}$ of $x\in\F_{2^n}$, where $n=2m$; 
\item $U_m$: the unit circle $$ U_m=\{z\in\F_{2^n}:z^{2^m+1}=1\}, $$ where $n=2m$; 
\item $\Tr_m^n(\cdot)$: the trace function from $\F_{2^n}$ to $\F_{2^m}$, where $m\mid n$, defined by $$ \Tr_m^n(x)=x+x^{2^m}+\cdots+x^{2^{(\frac{n}{m}-1)m}}. $$ When $m=1$, it becomes the absolute trace function and is denoted by $\Tr_n(\cdot)$ for simplicity.

We adopt the convention $0^{-1}=0$.
Moreover, whenever an expression $f(x)$ is regarded as a function
on $\F_{2^n}$, every nonzero integer exponent $a$ occurring in a
monomial $x^a$ is understood modulo $2^n-1$. If
$a\equiv0\pmod{2^n-1}$, the exponent is represented by $2^n-1$
rather than by $0$.

\end{itemize} 

\subsection{Boolean Functions and Walsh Transform}
Let $n$ and $m$ be positive integers. An $(n,m)$-function is a mapping from $\mathbb F_{2^n}$ to $\mathbb F_{2^m}$. In particular, when $m = 1$, such a mapping is referred to as a Boolean function over $\mathbb{F}_{2^n}$. Several canonical representations exist for Boolean functions.

One fundamental representation is the algebraic normal form (ANF), which provides a unique polynomial expression over $\mathbb{F}_{2^n}$. Specifically, any Boolean function $f\colon \mathbb{F}_{2^n} \to \mathbb{F}_2$ admits a unique expansion  
$$
f(x) = \sum_{i=0}^{2^n - 1} c_i x^i,
$$
where the coefficients $c_i$ satisfy the following constraints: $c_0, c_{2^n - 1} \in \mathbb{F}_2$, and for $1 \leq i < 2^n - 1$, the condition $c_{2i \bmod (2^n - 1)} = c_i^2$ holds.

Another widely adopted representation is the trace representation. A Boolean function $f$ over $\mathbb{F}_{2^n}$ can be expressed as  
$$
f(x) = \operatorname{Tr}_n(H(x)),
$$
where $H$ is an $(n,n)$-function. Unlike the ANF, this representation is generally non-unique.

Finally, a Boolean function may be  characterized by its support. For a Boolean function $f\colon \mathbb{F}_{2^n} \to \mathbb{F}_2$, the support is defined as the subset $\operatorname{Supp}(f) = \{x \in \mathbb{F}_{2^n} \mid f(x) = 1\}$. Conversely, given any subset $S \subseteq \mathbb{F}_{2^n}$, the indicator function of $S$—denoted $f_S$—is the unique Boolean function over $\mathbb{F}_{2^n}$ satisfying  
$$
f_S(x) =
\begin{cases}
1, & \text{if } x \in S, \\
0, & \text{if } x \notin S.
\end{cases}
$$

For a Boolean function $f$ over $\F_{2^n},$ its Walsh transform at $\gamma\in\F_{2^n}$ is defined as
$$W_f(\gamma)=\sum_{x\in \F_{2^n}}(-1)^{f(x)+\Tr_n(\gamma x)}=\sum_{x\in \F_{2^n}}(-1)^{f(x)}\chi_n(\gamma x),$$
where $\chi_n(z)=(-1)^{\Tr_n(z)}$ denotes the canonical additive character on $\F_{2^n}.$

The Walsh transform satisfies several fundamental identities.  
The first is the inverse Walsh transform formula. Indeed, the Walsh transform can be viewed as the Fourier transform of the sign function
$x\mapsto (-1)^{f(x)}$ on the additive group of $\mathbb F_{2^n}$.  Consequently, applying the standard Fourier inversion theorem yields  
$$
(-1)^{f(x)}
=
2^{-n}\sum_{\gamma\in\mathbb F_{2^n}}
W_{f}(\gamma)\chi_n(\gamma x).
$$

This Fourier interpretation also gives Parseval's identity. Namely, applying Parseval's formula to the sign function $x\mapsto (-1)^{f(x)}$, we have
$$
\sum_{\gamma\in\mathbb F_{2^n}} W_{f}(\gamma)^2
=
2^n\sum_{x\in\mathbb F_{2^n}} |(-1)^{f(x)}|^2=2^{2n}.
$$

These identities play a central role in analyzing the Walsh value distribution of a Boolean function $f\colon \mathbb{F}_{2^n} \to \mathbb{F}_2$. Moreover, suppose its Walsh values take only the $t$ distinct values $a_1,\dots,a_t$. We define the Walsh spectrum of $f$ as 
$
\{w_{a_1},\dots,w_{a_t}\},$
where
$
w_{a_i}:=\#\{x\in\mathbb F_{2^n}: W_f(x)=a_i\}.
$
Then we have the following identities:
\begin{equation}\label{eq2.1}
\begin{cases}
    \sum_{i=1}^t w_{a_i}=2^n,\\
    \sum_{i=1}^t w_{a_i} a_i=2^n (-1)^{f(0)},\\
    \sum_{i=1}^t w_{a_i}a_i^2=2^{2n}.
\end{cases}
\end{equation}
Here the first identity follows from the definition of $M_i$, while the second
and the third ones follow from the inverse Walsh transform formula at $x=0$ and
Parseval's identity, respectively.

At the end of this subsection, we introduce the Walsh transform for general $(n,n)$-functions. For a $(n,n)$-function $F,$ its Walsh transform at $(u,v)\in\F_{2^n}\times \F_{2^n}$ is defined as follows:
$$W_F(u,v)=\sum_{x\in\F_{2^n}}(-1)^{\Tr_n(vF(x))+\Tr_n(ux)}=\sum_{x\in\F_{2^n}}\chi_n(vF(x)+ux).$$

\subsection{Cryptographic criteria of Boolean functions}
In this subsection, we recall several cryptographic properties of Boolean
functions that will be used in this paper. In particular, we focus on three basic parameters: balancedness, algebraic degree, and nonlinearity. These
parameters measure different aspects of the cryptographic strength of Boolean functions and are intrinsically linked to both the Walsh spectrum and the algebraic normal form (ANF) of the function.

To avoid statistical dependence between the plaintext and ciphertext, Boolean functions used in cryptographic applications
are usually required to be balanced.

\begin{definition}
    A Boolean function $f$ is called balanced  if $\wt(f)=\wt(f+1)=2^{n-1},$ where $\wt(f)=\#\{x\in \F_{2^n}: f(x)=1\}.$ 
\end{definition}
This condition can be characterized by the Walsh transform. Indeed, since
$$
W_f(0)=\sum_{x\in\mathbb F_{2^n}}(-1)^{f(x)}
=\#\{x:f(x)=0\}-\#\{x:f(x)=1\},
$$
we immediately obtain the following criterion.
\begin{proposition}
    A Boolean function $f$ is balanced if and only if $W_f(0)=0.$
\end{proposition}

To withstand the Berlekamp-Massey and the Rønjom-Helleseth attack, a high algebraic degree is needed. The algebraic degree, denoted by $\deg(f)$ of a Boolean function $f$ over $\F_{2^n}$ can be defined from its univariate representation.
\begin{definition}
    Let $f$ be a Boolean function over $\F_{2^n}.$  If $f(x) = \sum_{i=0}^{2^n - 1} c_i x^i$ denotes its unique algebraic normal form, then the algebraic degree of $f$, denoted $\deg(f)$, is defined as  
$$
\deg(f) = \max_{0\leq i\leq 2^n-1}\big\{ \operatorname{wt}(i) \mid c_i \neq 0 \big\},
$$  
where $\operatorname{wt}(i)$ denotes the Hamming weight of the binary expansion of the integer $i$.
\end{definition}

In fact, the algebraic degree of a Boolean function $f$ equals $n$ if and only if its Hamming weight $\operatorname{wt}(f)$ is odd. Consequently, the maximum possible algebraic degree of a balanced Boolean function over $\mathbb{F}_{2^n}$ is $n-1$.

To resist the best affine approximation and the fast correlation attack, Boolean functions used in cryptography must have high nonlinearity.
\begin{definition}
    For a Boolean function over $\F_{2^n},$ its nonlinearity is defined as $\mathcal{N}\mathcal{L}(f)=\min_{g\in \mathcal{A}_n}\wt(f+g),$ where $\mathcal{A}_n$ is the set of all affine Boolean function over $\F_{2^n},$ whose algebraic degrees are at most $1.$
\end{definition}

The nonlinearity can also be determined by the Walsh transform.
\begin{proposition}[\cite{carlet2021boolean}]\label{prop4}
    For a Boolean function $f$ over $\F_{2^n},$
    $$\mathcal{N}\mathcal{L}(f)=2^{n-1}-\frac{1}{2}\max_{a\in\F_{2^n}}|W_f(a)|.$$
\end{proposition}

By the parseval's identity, we obtain $\max_{a\in \F_{2^n}}|W_f(a)|\geq 2^{\frac{n}{2}}.$ Consequently, the nonlinearity of $f$ satisfies $\mathcal{N}\mathcal{L}(f)\leq 2^{n-1}-2^{\frac{n}{2}-1}.$ This upper bound is attained if and only if
$\max_{a\in\mathbb F_{2^n}} |W_f(a)|=2^{\frac n2}.$
By Parseval's identity again, this happens if and only if
$W_f(a)\in \left\{\pm 2^{\frac n2}\right\}$
for every $a\in\mathbb F_{2^n}$. In this case, $f$ is called a bent function. 

Clearly, bent functions can exist only when $n$ is even. Moreover, they are not
balanced, since $W_f(0)\in \left\{\pm 2^{\frac n2}\right\}.$
To allow Walsh spectra with zero values and to include broader classes of
cryptographically interesting functions, we introduce plateaued functions.
\begin{definition}
Let $n$ and $k$ be positive integers with $n\equiv k \pmod 2$. A Boolean
function $f$ over $\mathbb F_{2^n}$ is called a $k$-plateaued function if
$
W_f(a)\in \left\{0,\pm 2^{\frac{n+k}{2}}\right\}
$
for every $a\in\mathbb F_{2^n}$.
\end{definition}

Finally, to evaluate the resistance of Boolean functions against algebraic attacks, we also recall the concept of algebraic immunity.
\begin{definition}
   Let $f$ and $h$ be Boolean functions over $\mathbb F_{2^n}.$ We say that $h$ is an annihilator of $f$ if $fh=0.$ The algebraic immunity $\operatorname{AI}(f)$ of $f$ is defined as the minimum algebraic degree of a nonzero Boolean function $h$ such that $h$ is annihilator of $f$ or $f+1.$
\end{definition}

\begin{proposition}[\cite{courtois2003algebraic}]
   For a Boolean function $f$ over $\F_{2^n},$ 
   $$\operatorname{AI}(f)\leq \lceil\frac{n}{2}\rceil.$$
\end{proposition}

Algebraic immunity is expected to be large in order to resist algebraic attacks. A Boolean function $f$ is said to have optimal algebraic immunity if $\operatorname{AI}=\lceil\frac{n}{2}\rceil,$ and near-optimal algebraic immunity if $\operatorname{AI}=\lceil\frac{n}{2}\rceil-1.$

\subsection{Boolean functions from $2$-to-$1$ mappings}
Let $F$ be a $2$-to-$1$ mapping from $\mathbb F_{2^n}$ to itself; that is, for each $a\in\mathbb F_{2^n}$, the preimage $F^{-1}(a)$ is either empty or consists of exactly two elements. Using the 2-to-1 property of $F,$ we can construct a balanced Boolean function $f,$ whose support equals the image set of $F,$ i.e,
$$f(x)=\begin{cases}
    1, &x\in \im(F),\\
    0, &x\not\in \im(F).
\end{cases}$$
This construction is referred to as the parametric construction, and we say that $f$ is parameterized by $F.$ In particular, the induced mapping of $x^2+x$ and $x+x^{-1}$ is 2-to-1 over $\F_{2^n}.$ Furthermore, for any permutation polynomial $P(x) \in \mathbb{F}_{2^n}[x]$, the compositions $P(x^2 + x)$ and $P(x + x^{-1})$ also define $2$-to-$1$ mappings on $\mathbb{F}_{2^n}$. For $2$-to-$1$ mappings of the form $P(x^2 + x)$ and $P(x + x^{-1})$, Qu, Zhang, and Li \cite{qu2025parametric} established an explicit relationship between the Walsh transform of a Boolean function parameterized by such a mapping and the underlying permutation polynomial $P(x)$.
\begin{lemma}[\cite{qu2025parametric}]\label{lem1}  
    Let $P$ be a permutation over $\F_{2^n}.$  Then we have the following results.
    \begin{enumerate}
        \item [(i)] If $f$ is a balanced  Boolean function  parameterized by $P(x+x^2).$ Then for any $a\in \F_{2^n}^*,$
    $$W_f(a)=-\sum_{x\in\F_{2^n}}\chi_n(aP(x)+x)=-W_P(1,a).$$
    \item [(ii)] If $f$ is a balanced Boolean function  parameterized by $P(x+x^{-1}).$ Then for any $a\in \F_{2^n}^*,$
    $$W_f(a)=-\sum_{x\in\F_{2^n}}\chi_n(aP(x)+x^{-1}).$$
    \end{enumerate}
\end{lemma}

\begin{lemma}[\cite{zhang2026balanced}]\label{lemw1}
    Let $n=2m$ and $G(x)=x^rh(x^{2^m-1}),$ where $r$ is a positive integer and $h(x)$ is a polynomial over $\F_{2^n}.$ Then the Walsh transform of $G$ at $(a,b)\in \F_{2^n}^2$ is given by
    $$W_G(a,b)=-2^m+\sum_{z\in U_m}\sum_{y\in \F_{2^m}}\chi_m(\Tr_m^n(bz^rh(\bar{z}^2))y^r+\Tr_m^n(az)y).$$
\end{lemma}

In particular, for the case $r=1$, Lemma \ref{lemw1} can reduce the computation of $W_G(a,b)$ to the number of roots in the unit circle $U_m$ of a certain equation. More precisely, we have the following result.
\begin{lemma}[\cite{zhang2026balanced}]\label{lemwa}
    Let $n=2m$ and $P(x)=xh(x^{2^m-1})\in \F_{2^n}\left[x\right]$ is a permutation polynomial, where $h(x)$ is a polynomial over $\F_{2^n}.$ If Boolean function $f$ is parameterized by $P(x+x^2),$ then for any $\gamma\in\F_{2^n}^*,$
    $$W_f(\gamma)=-2^m\left(\# \{z\in U_m: \Tr_m^n\big(\gamma zh(\bar{z}^2)\big)+z+\bar{z}=0\}-1\right).$$
\end{lemma}

\subsection{Permutation polynomials from Niho exponents}

Let $q=2^m$. An integer $d$ is called a Niho exponent with respect to $\mathbb F_{q^2}$ if $d\equiv 2^j \pmod {q-1}$ for some nonnegative integer $j$. In particular, when $j=0$, $d$ is called a normalized Niho exponent. More generally, an integer $d$ is called a generalized Niho exponent if $d\equiv \ell \pmod {q-1}$ for some fixed integer $\ell$, where $\ell$ is not necessarily a power of $2$. Niho exponents and generalized Niho exponents have found applications in several areas, including coding theory, bent functions, and permutation polynomials. For more details, we refer the reader to \cite{li2019survey}.

In this paper, we primarily focus on the balanced Boolean function parameterized by $P(x+x^2)$ and $P(x+x^{-1}),$ where $P(x)$ is a permutation polynomial of the form $x^r h(x^{q-1})\in \F_{q^2}\left[x\right].$ As all monomial exponents appearing in $P(x)$ are congruent to $r$ modulo $q-1$, they constitute a common congruence class of (generalized) Niho exponents.  

In the following, we give three known classes of permutation polynomials over $\mathbb F_{q^2}.$

\begin{proposition}[\cite{ding2025class}]\label{prop1}
 Let $m$ be a positive integer. Then the polynomial $x(1+x^{2^m-1}+x^{2(2^m-1)})$ permutes $\F_{2^{2m}}$ if and only if $m$ is even.    
\end{proposition}

\begin{proposition}[\cite{gupta2016some}]\label{prop2}
  Let $m$ be a positive integer. Then the polynomial $x^2(1+x^{2(2^m-1)}+x^{3(2^m-1)})$ permutes $\F_{2^{2m}}$ if and only if $\gcd(m,3)=1.$
\end{proposition}

\begin{proposition}[\cite{ding2015permutation}]\label{prop3}
    Let $m$ and $k$ be  positive integers. Then the polynomials $x(1+x^{k(2^m-1)}+x^{-k(2^m-1)})$ permutes $\F_{2^{2m}}$ if and only if  $m$ is even.
\end{proposition}

Finally, we recall two elementary lemmas that are frequently used when dealing with Niho-type exponents. The first one is the polar decomposition of $\mathbb F_{q^2}^*$, and the second one provides a parametrization of the unit circle $U_m\setminus\{1\}$ by $\F_{2^m}$. these lemmas provide a finite reduction from problems over $\mathbb F_{q^2}$ to problems over $\mathbb F_q\times\F_q.$

\begin{lemma}[\cite{dobbertin2006niho}]
    Let $n=2m.$ Then for any $x\in \F_{2^n}^*,$ there exists a unique pair $(y,z)\in \F_{2^m}^*\times U_m$ such that $x=yz.$
\end{lemma}

\begin{lemma}\label{lem3.2.3}
    Let $n=2m$ and $q=2^m.$ Let $a$ be an element of $\F_q$ such that
$$\begin{cases}
    \Tr_m(a)=0, & \text{ if } m \text{ is odd};\\
    \Tr_m(a)=1, & \text{ if } m \text{ is even}.\\
\end{cases}$$
Then there exists a bijection  between $\F_{q}$ and $U_m\setminus\{1\},$ which is given by
\begin{align*}
    \varphi: \F_q&\rightarrow U_m\setminus\{1\}\\
    x&\mapsto \frac{x+\theta}{\sqrt{x^2+x+a+1}},
\end{align*}
where $\theta$ is a root of $x^2+x+a+1=0.$
\end{lemma}
\begin{proof}
    Let $D(x)=x^2+x+a+1.$ By the choice of $a,$ we have $\Tr_m(a+1)=1.$ It follows that $D(x)$ is irreducible over $\F_q$. Therefore, $\theta\in\F_{q^2}\setminus\F_q$, and $D(x)\neq 0$ for every $x\in\F_q$.

    Since the other root of $x^2+x+a+1$ is $\theta+1$, we obtain
    $$ D(x)=(x+\theta)(x+\theta+1)=(x+\theta)(x+\theta^q). $$
    Consequently, $ \theta^q=\theta+1 $ and
    $$ \left(\frac{x+\theta}{\sqrt{D(x)}}\right)^{q+1} = \frac{x+\theta}{\sqrt{D(x)}}\cdot \frac{x+\theta^q}{\sqrt{D(x)}} = \frac{(x+\theta)(x+\theta+1)}{D(x)} =1. $$ Hence, $\varphi(x)\in U_m$. Moreover, $\varphi(x)\neq 1$, since otherwise $x+\theta=\sqrt{D(x)}\in\F_q$, which would imply $\theta\in\F_q$, a contradiction. Therefore, $\varphi$ is well defined as a map from $\F_q$ to $U_m\setminus\{1\}$.

Since $\#\F_q=\#U_m\setminus\{1\},$ in order to show $\varphi$ is a bijection, it suffices to prove that $\varphi$ is injective. Let $x_1, x_2$ be two distinct elements of $\F_q$ such that
\begin{equation*}
    \frac{x_1+\theta}{\sqrt{x_1^2+x_1+a+1}}=\frac{x_2+\theta}{\sqrt{x_2^2+x_2+a+1}}.
\end{equation*}
Equivalently, 
$$(x_1+x_2)\left(x_1x_2+(x_1+x_2)(a+1)+\theta^2(x_1+x_2+1)\right)=0.$$
The assumption $x_1\ne x_2$ implies that
\begin{equation}\label{eq24}
\left(x_1x_2+(x_1+x_2)(a+1)+\theta^2(x_1+x_2+1)\right)=0.    
\end{equation}
Note that $1+x_1+x_2\ne0$. Indeed, if $1+x_1+x_2=0$, then \eqref{eq24} would reduce to
$$
D(x_1)=x_1^2+x_1+a+1=0,
$$
contradicting the irreducibility of $D(x)$ over $\mathbb F_q$.

which implies that $\theta^2=\frac{x_1x_2+(x_1+x_2)(a+1)}{1+x_1+x_2}\in \F_q,$ which leads to a contradiction.  This completes the proof. \hfill\ensuremath{\mbox{$\Box$}}
\end{proof}
\section{Constructions of few-valued spectrum Boolean function}\label{sec3}
In this section, based on the permutation polynomials from  Proposition \ref{prop1} and \ref{prop2},  we construct two balanced Boolean functions parameterized by $P(x+x^2)$ with few-valued Walsh spectra.

\subsection{Four-valued spectrum}
Using the permutation polynomial from Proposition \ref{prop1}, we first construct a balanced Boolean function parameterized by $Q(x+x^2)$ and show that it has a four-valued Walsh spectrum.
\begin{theorem}\label{thm1}
    Let $q=2^m$ with $m$ even, $Q(x)=xh(x^{q-1})\in \F_{q^2}\left[x\right],$ where $h(x)=1+x+x^2,$ and $f_Q$ be the balanced Boolean function parameterized by $Q(x+x^2).$ Then the Walsh spectrum of $f_Q$ is
$$\{w_q, w_0, w_{-q}, w_{-2q}\},$$
where 
$$w_q=\frac{q^2-q}{3}, ~w_0=\frac{q^2}{2}-\frac{q}{2}, ~w_{-q}=q,~ w_{-2q}=\frac{q(q-1)}{6}.$$
\end{theorem}

\begin{proof}
     For $\gamma=0,$ the balancedness of $f_Q$ implies that $W_{f_Q}(0)=0.$ For any given $\gamma\in \F_{q^2}^*$, by Lemma~\ref{lemwa}, we obtain that
\begin{equation}\label{eq1}
    W_{f_Q}(\gamma)=-q\left(\# \{z\in U_m: \text{Tr}_m^{2m}\big(\gamma zh(\bar{z}^2)\big)+z+\bar{z}=0\}-1\right).
\end{equation}
Since $h(x)=1+x+x^2,$ we can directly compute that
$$\text{Tr}_m^{2m}\big(\gamma zh(\bar{z}^2)\big)+z+\bar{z}=\bar{\gamma}z^3+(1+\gamma+\bar{\gamma})(z+\bar{z})+\gamma \bar{z}^3.$$
For any $z\in U_m,$ we have $z\bar{z}=1.$ Then
$$\bar{\gamma}z^3+(1+\gamma+\bar{\gamma})(z+\bar{z})+\gamma \bar{z}^3=0\Longleftrightarrow \bar{\gamma}z^6+(1+\gamma+\bar{\gamma})(z^2+z^4)+\gamma=0.$$
Since the map $x\mapsto x^2$ permutes $U_m,$ we have the number of roots of $\bar\gamma z^6+(1+\gamma+\bar{\gamma})(z^2+z^4)+\gamma$ equal to the number of roots of $\bar{\gamma}z^3+(1+\gamma+\bar{\gamma})(z+z^2)+\gamma=0.$ Therefore, for any $\gamma\in \F_{q^2}^*,$ by (\ref{eq1}), we have
\begin{equation}\label{eq2}
     W_{f_Q}(\gamma)=-q\left(\# \{z\in U_m: z^3+\frac{(1+\gamma+\bar{\gamma})}{\bar{\gamma}}(z+z^2)+\frac{\gamma}{\bar{\gamma}}=0\}-1\right).
\end{equation}
For any $\gamma\in \F_{q^2}^*,$  write $\mathcal{P}_\gamma(z)=z^3+\frac{(1+\gamma+\bar{\gamma})}{\bar{\gamma}}(z+z^2)+\frac{\gamma}{\bar{\gamma}}.$ The number of roots of $\mathcal{P}_\gamma(z)$ in $U_m$ can only be $0,1,2,$ or $3.$ Then by (\ref{eq2}) and $W_{f_Q}(0)=0$, we have
$$W_{f_Q}(\gamma)\in\{2^m,0,-2^m,-2^{m+1}\}.$$

 Since $f_Q(0)=1,$ by (\ref{eq2.1}), we have

\begin{equation}\label{eq3}
\begin{cases}
    w_{q}+w_{0}+w_{-q}+w_{-2q}=q^2,\\
    w_{q}\cdot q+ w_0\cdot 0+w_{-q} \cdot(-q)+w_{-2q}\cdot (-2q)=-q^2,\\
    w_{q}\cdot q^2 +w_0 \cdot 0^2+w_{-q} \cdot q^2+w_{-2q}\cdot 4q^2=q^4.
\end{cases}    
\end{equation}
Solving the linear system and taking $w_{-q}$ as a free variable, we obtain 
\begin{equation}\label{eq4}
\begin{cases}
w_{q}=\dfrac{w_{-q}+q^2-2q}{3},\\
w_0=\frac{q^2}{2}+\frac{q}{2}-w_{-q},\\
w_{-2q}=\dfrac{q^2+q-2w_{-q}}{6}.
\end{cases}    
\end{equation}

We now determine $w_{-q}.$ Let $\mathcal{W}$ consist of all $\gamma\in\F_{q^2}^*$ for which $\mathcal{P}_\gamma(z)$ has exactly two distinct roots in $U_m$. By the definition of $w_{-q}$ and $W_{f_Q}(0)=0,$  we have  $w_{-q}=\#\mathcal{W}.$ Let $z_0,z_1,z_2$ be the roots of
$\mathcal{P}_\gamma(z)$ in some algebraic closure $\overline{\F}_2$.
By Vieta's formula, we have
$z_0z_1z_2=\frac{\gamma}{\bar{\gamma}}.$ Since $\frac{\gamma}{\bar{\gamma}}\in U_m,$ we obtain that if $z_0,z_1\in U_m,$ we have $z_2\in U_m.$  Therefore, $\mathcal{P}_\gamma(z)$ has exactly two distinct roots in $U_m$ if and only if $\mathcal{P}_\gamma(z)=(z+z_0)^2(z+z_1),$ where $z_0,z_1$ are two distinct elements of $U_m.$ In this case, since the coefficients of $z^2$ and $z$ in $\mathcal{P}_\gamma(z)$ are equal, expanding $(z+z_0)^2(z+z_1)$ and comparing coefficients gives
\[
z_0+z_0+z_1 = z_0^2+z_0z_1+z_0z_1 = \frac{1+\gamma+\bar\gamma}{\bar\gamma},
\]
it follows that $z_1=z_0^2.$ Note that $z_0^2\ne z_0$ if and only if $z_0\in U_m\setminus \{1\}.$ Thus $\mathcal{P}_\gamma(z)$ has exactly two distinct roots in $U_m$ if and only if $$\mathcal{P}_\gamma(z)=(z+z_0)^2(z+z_0^2),$$ where $z_0\in U_m\setminus\{1\}.$ In this case, compare the coefficients of $\mathcal{P}_\gamma(z),$ we obtain 
$$\frac{1+\gamma+\bar{\gamma}}{\bar{\gamma}}=z_0^2, ~ ~\frac{\gamma}{\bar{\gamma}}=z_0^4.$$
Since $m$ is even, $U_m$ contains no nontrivial cube roots of unity. It follows that $1+z_0^2+z_0^4\ne 0$ for each $z_0\in U_m\setminus\{1\}.$ Then we have 
$$\gamma=\frac{z_0^4}{1+z_0^2+z_0^4}, ~~\bar{\gamma}=\frac{1}{1+z_0^2+z_0^4}.$$ Therefore, the map
\begin{align*}
    U_m\setminus\{1\}&\longrightarrow \mathcal{W},\\
    z_0&\longmapsto\frac{z_0^4}{1+z_0^2+z_0^4}
\end{align*}
is a bijection.
Hence $w_{-q}=q$. Substituting this into \eqref{eq4} yields the Walsh spectrum of $f_Q$ as stated. \hfill\ensuremath{\mbox{$\Box$}}
\end{proof}

By Proposition \ref{prop4}, we obtain the non-linearity of $f_Q.$ 
\begin{corollary}
 Let $q=2^m$ with $m$ even, and let $f_Q$ be the balanced Boolean function defined in Theorem \ref{thm1}. Then
$$
\mathcal{N}\mathcal{L}(f_Q)=\frac{q^2}{2}-q.
$$
\end{corollary}

\begin{example}
For further illustration, we give several explicit examples of the Boolean function $f_Q$ obtained in Theorem \ref{thm1} for $m=2,4,6 ,n=2m$. The corresponding trace representations, Walsh spectral distributions, nonlinearity, algebraic degree, and algebraic immunity are listed in Table \ref{tab:four-valued}. Here, the notation $a^b$ means that the Walsh value $a$ occurs $b$ times in the Walsh spectrum.
\begin{table}[htbp]
\centering
\caption{Four-valued Walsh spectrum Boolean functions defined in Theorem \ref{thm1}}\label{tab:four-valued}
\begin{threeparttable}
\begin{tabular}{cccccc}
\toprule
$n$ & $f_Q(x)=\operatorname{Tr}_n(R_n(x))+1$ & Walsh spectrum & $\mathcal{N}\mathcal{L}(f_Q)$ & $\deg(f_Q)$ & $\mathrm{AI}(f_Q)$ \\
\midrule
 4 & $R_4(x)=x^7+x$ & $\{0^6, 4^4, -4^4, -8^2\}$ &  4& 3 &  2\\
 8 & $R_8(x)$\tnote{1} & $\{0^{120}, -32^{40}, 16^{80}, -16^{16}\}$ &  112&  5 &  4\\
 12 & $R_{12}(x)$\tnote{2}  & $\{0^{2016}, 64^{1344}, -128^{672}, -64^{64}\}$ & 1984 & 7 & 5 \\
\bottomrule
\end{tabular}
\begin{tablenotes}[flushleft]
\footnotesize
\item[1] $R_8(x)=x^{61} + x^{53} + x^{47} + x^{31} + x^{19} + x.$

\item[2] $R_{12}(x)=x^{1387} + x^{949} + x^{827} + x^{823} + x^{725} + x^{695} + x^{631} + x^{599} + x^{505} + x^{457} + x^{347} + x^{319} + x^{317} + x^{253} + x^{221}
+ x^{205} + x^{197} + x^{191} + x^{127} + x^{79} + x^{67} + x.$

\end{tablenotes}
\end{threeparttable}
\end{table}

As shown in Table \ref{tab:four-valued}, although the univariate trace representation of $f_Q$ becomes increasingly complicated as $n$ grows, its Walsh spectrum remains completely determined by Theorem \ref{thm1}. In particular, these examples confirm that the constructed Boolean functions have high nonlinearity $\mathcal{N}\mathcal{L}(f_Q)=2^{n-1}-2^m.$
Moreover, for $n \leq 12$, the numerical results also indicate that the algebraic immunity of these functions is very close to the optimal upper bound, and their algebraic degree equals $m+1$.

\end{example}

\subsection{A parameterized family with a five-valued Walsh spectrum}
In this subsection, we construct a parameterized family of
permutation polynomials $P_\lambda$ and investigate the balanced
Boolean functions parameterized by $P_\lambda(x+x^2)$. We first
characterize and enumerate the admissible parameters $\lambda$, and
then prove that all the resulting Boolean functions share the same
five-valued Walsh distribution. We begin with several auxiliary
results on root counting for cubic and quartic polynomials.

\begin{lemma}[\cite{berlekamp1967solution}]\label{lem3.2.1}
Let $a,b\in \F_{2^n},$ where $b\ne0.$ Then the cubic equation $x^3+ax+b=0$ has a unique solution in $\F_{2^n}$ if and only if $\text{Tr}_n(\frac{a^3}{b^2}+1)\ne 0.$
\end{lemma}

In the following lemma, the notation $f=(d_1,d_2,\ldots,d_s)$ means that $f$ factors over the underlying field  into a product of  irreducible polynomials whose degrees are precisely $d_1, d_2, \dots, d_s$, respectively. 

\begin{lemma}[\cite{leonard1972quartics}]\label{lem3.2.2}
	Let $f(x)=x^4+a_2x^2+a_1x+a_0$ with $a_i\in \mathbb{F}_{2^n}$ and $a_0a_1\neq 0$. Let $f_1(y)=y^3+a_2y+a_1$ and $r_1,r_2,r_3$ denote roots of $f_1(y)=0$ when they exist in $\mathbb{F}_{2^n}$. Set $w_i=a_0\dfrac{r_i^2}{a_1^2}$. Then the factorization of $f(x)$ over $\mathbb{F}_{2^n}$ is characterized as follows:
	
1. $f=(1,1,1,1)$ if and only if $f_1=(1,1,1)$ and $\operatorname{Tr}_{2^n}(w_1)=\operatorname{Tr}_{2^n}(w_2)=\operatorname{Tr}_{2^n}(w_3)=0$;
		
2. $f=(2,2)$ if and only if $f_1=(1,1,1)$ and $\operatorname{Tr}_{2^n}(w_1)=0$, $\operatorname{Tr}_{2^n}(w_2)=\operatorname{Tr}_{2^n}(w_3)=1$;
		
3. $f=(1,3)$ if and only if $f_1=(3)$;
		
4. $f=(1,1,2)$ if and only if $f_1=(1,2)$ and $\operatorname{Tr}_{2^n}(w_1)=0$;
		
5. $f=(4)$ if and only if $f_1=(1,2)$ and $\operatorname{Tr}_{2^n}(w_1)=1$.
\end{lemma}

\begin{lemma}\label{lem3.2.4}
    Let $n=2m, a\in\F_{2^n}^*$ and $b\in U_m.$ Then the polynomial $f(x)=x^4+a(x^3+x)+b$ has exactly three distinct roots in $U_m$ if and only if $f(x)=(x+1)^2(x+z)(x+z^{-1}),$ where $z\in U_m\setminus\{1\}.$
\end{lemma}

\begin{proof} The sufficiency is clear. It remains to prove
the necessity. Suppose that $f(x)$ has exactly three distinct roots in $U_m.$ Let these roots be $z_0,z_1,z_2\in U_m$, and let $y$ be the fourth root of $f(x)$ in some algebraic closure.  By Vieta's formula, we have $z_0z_1z_2y=b\in U_m.$ It follows that $y\in U_m.$ Since $f(x)$ has exactly three distinct roots in $U_m$, we have
$y\in\{z_0,z_1,z_2\}$. Without loss of generality, assume that $y=z_0$. Therefore, we have $f(x)=(x+z_0)^2(x+z_1)(x+z_2).$
Comparing the coefficient of $x^3, x^2$ and $x$, we can deduce that
\begin{equation}\label{eq5}
z_1+z_2=z_0^2z_1+z_0^2z_2=a,    
\end{equation}
and 
\begin{equation}\label{eq6}
z_0^2+z_1z_2=0.
\end{equation}
Then by (\ref{eq5}) and $z_1\ne z_2,$ we have $z_0=1.$ Substituting $z_0=1$ into (\ref{eq6}), we get $z_2=z_1^{-1}.$ Hence, 
$$f(x)=(x+1)^2(x+z_1)(x+z_1^{-1})$$
as desired. \hfill\ensuremath{\mbox{$\Box$}}
\end{proof}

Comparing the constant terms in the factorization above, we
immediately obtain the following corollary.

\begin{corollary}\label{cor3.2.1}
    Let $n=2m, a\in\F_{2^n}^*$ and $b\in U_m.$ If the polynomial $f(x)=x^4+a(x^3+x)+b$ has exactly three distinct roots in $U_m,$ then $b=1.$ 
\end{corollary}

Gupta and Sharma~\cite{gupta2016some} investigated the permutation
property of the polynomial $f_1(x)=x^2\left(1+x^{2(q-1)}+x^{3(q-1)}\right)$ over $\F_{q^2}$. Motivated by their work, we extend this polynomial
to a parameterized family and obtain further permutation polynomials.
We also determine the complete Walsh distributions of the associated
balanced Boolean functions. We begin with a characterization of the
parameters for which the resulting polynomials are permutations.

\begin{theorem}[Permutation criterion]\label{thm:PlambdaPP}
Let $q=2^m$ and, for $\lambda\in\F_q$, define
\[
P_\lambda(x)=x^{3q-1}+\lambda x^{2q}+\lambda x^2
=x^2\bigl((x^{q-1})^3+\lambda(x^{q-1})^2+\lambda\bigr).
\]
Then $P_\lambda$ permutes $\F_{q^2}$ if and only if either $\lambda=0$, or
$\lambda\ne0$ and the polynomial
$X^3+X+\lambda^{-1}$ has no root in $\F_q$.
\end{theorem}

\begin{proof}
Put $h_\lambda(z)=z^3+\lambda z^2+\lambda$. By the multiplicative case of the AGW criterion \cite{lee1997some}, $P_\lambda(x)$ permutes $\mathbb F_{q^2}$ if and only if $g_\lambda(z):=z^2h_\lambda(z)^{q-1}$ permutes $U_m$. 

Notice first that $h_\lambda$ has no root in $U_m$. Indeed, this is clear when $\lambda=0$. If $\lambda\ne0$ and $h_\lambda(z)=0$, then $\lambda=\frac{z^3}{z^2+1}$. Taking the $q$-th power gives $\lambda=\frac{z^{-1}}{z^2+1}$, and hence $z^4=1$. Since $|U_m|=q+1$ is odd, this forces $z=1$, contradicting $h_\lambda(1)=1$.

Since $g_\lambda(1)=1$, it remains to show that $g_\lambda$ permutes $U_m\setminus\{1\}$. First, we claim that $g_\lambda(U_m\setminus\{1\})\subseteq U_m\setminus\{1\}$. To see this, it suffices to show that $g_\lambda(z)\ne1$ for every $z\in U_m\setminus\{1\}$. Suppose otherwise. Then a direct calculation gives
\[
z^{-1}+\lambda+\lambda z^2=z^3+\lambda z^2+\lambda,
\]
which is equivalent to $z^4=1$. Hence $z=1$, contradicting the assumption. This proves the claim.

For each $z\in U_m\setminus\{1\}$, the two elements $z$ and $\bar z$ are precisely the roots of  $x^2+(z+\bar{z})x+1=0.$ Define $u(z):=1/(z+\bar z)$. By the Artin–Schreier criterion, we have $\operatorname{Tr}_m(u(z))=1.$
$$T:=\{u: \Tr_m(u)=1\}.$$ Combining the above observations with the fact that $\frac{\#U_m\setminus\{1\}}{2}=\#T$, we obtain that  
the map 
\begin{align*}
    u : U_m\setminus\{1\}&\longrightarrow T\\
    z&\longmapsto u(z)
\end{align*}
is  two-to-one surjective. Since $g_\lambda(U_m\setminus\{1\})\subseteq U_m\setminus\{1\},$ the map $u(g_\lambda(z))$ is well defined. Therefore, combining with $g_\lambda(\bar{z})=\overline {g_\lambda(z)}$ , we can deduce that $g_\lambda$ permutes $U_m\setminus\{1\}$ if and only if the map $z\mapsto u(g_\lambda(z))$ is  surjective.

A direct calculation gives
\begin{equation}
u(g(z))
=u(z)^4+\lambda^2u(z)^2+\lambda u(z)=L_\lambda(u(z)),
\label{eq:glambda}
\end{equation}
where
$$L_\lambda(u):=u^4+\lambda^2u^2+\lambda u.$$
 One readily checks that $\Tr_m(L_\lambda (u))=\Tr_m(u).$ Consequently, $L_\lambda(T)\subseteq T$. So we may regard $L_\lambda$ as a map from $T$ to itself. 
Since $u: U_m\setminus\{1\}\to T$ is surjective, equation \eqref{eq:glambda} shows that $z\mapsto u(g_\lambda(z))$ is surjective if and only if $L_\lambda$ is bijective on $T$, i.e., $\ker(L_\lambda)=\{0\}$.

For $\lambda=0$, $L_0(u)=u^4$ is a permutation of $T$. For $\lambda\ne0$, we have $\ker(L_\lambda)=\{0\}$ if and only if 
\[
u^3+\lambda^2u+\lambda=0
\]
has no root in $\mathbb F_q$. Substituting $u=\lambda x^2$ , this is equivalent to 
\[
x^3+x+\lambda^{-1}=0
\]
having no root in $\mathbb F_q$. This completes the proof.
\hfill\ensuremath{\mbox{$\Box$}}
\end{proof}

Theorem~\ref{thm:PlambdaPP} characterizes the parameters $\lambda$
for which $P_\lambda$ is a permutation polynomial. To count these
admissible parameters, we recall the following special case of
Bluher's counting result. It implies that approximately one-third
of the elements of $\F_q$ are admissible.

\begin{lemma}[\cite{bluher2004on}]\label{lem:bluher-cubic-count}
Let $m\ge2$ and $q=2^m$. The number of elements
$c\in\F_q^*$ for which $X^3+X+c$ has no root in $\F_q$ is
$$
\begin{cases}
(q-1)/3,&m\text{ even},\\[1mm]
(q+1)/3,&m\text{ odd}.
\end{cases}
$$
\end{lemma}

Combining Theorem~\ref{thm:PlambdaPP} with
Lemma~\ref{lem:bluher-cubic-count}, and noting that
$\lambda\mapsto\lambda^{-1}$ is a bijection of $\F_q^*$,
we immediately obtain the following corollary.

\begin{corollary}\label{prop:lambdacount}
Let $m\ge2$, $q=2^m$, and define
$$
\Lambda_m=\{0\}\cup
\{\lambda\in\F_q^*:X^3+X+\lambda^{-1}
\text{ has no root in }\F_q\}.
$$
Then
$$
|\Lambda_m|=
\begin{cases}
(q+2)/3,&m\text{ even},\\[1mm]
(q+4)/3,&m\text{ odd}.
\end{cases}
$$
\end{corollary}

We now determine the complete Walsh distributions of the
balanced Boolean functions parameterized by
$P_\lambda(x+x^2)$. The following theorem shows that these
distributions are the same for all $\lambda\in\Lambda_m$.

\begin{theorem}\label{thm:lambdaWalsh}
Let $m\ge2$, $q=2^m$, and $\lambda\in\Lambda_m$. Let $f_\lambda$ denote the balanced Boolean function parameterized by $P_\lambda(x+x^2)$. 
Then the complete Walsh distribution of $f_\lambda$ is independent of $\lambda$, and its Walsh spectrum is given as follows.

If $m$ is odd, the spectrum is
\[
\{w_q,\ w_0,\ w_{-q},\ w_{-2q},\ w_{-3q}\},
\]
where
\[
w_q=3\cdot\frac{q}{4}\left(\frac{q}{2}-1\right),~
w_0=\frac{q(2q+5)}{6},~
w_{-q}=\frac{q}{2}\left(\frac{q}{2}-1\right),~
w_{-2q}=\frac{q}{2},~
w_{-3q}=\frac{q(q-2)}{24}.
\]

If $m$ is even, the spectrum is
\[
\{w_q,\ w_0,\ w_{-q},\ w_{-2q},\ w_{-3q}\},
\]
with
\[
w_q=\frac{q(3q-4)}{8},~
w_0=\frac{q(2q+1)}{6},~
w_{-q}=\frac{q^2}{4},~
w_{-2q}=\frac{q}{2},~
w_{-3q}=\frac{q(q-4)}{24}.
\]
\end{theorem}

\begin{proof}
Throughout the proof, we fix $\lambda\in\Lambda_m$ and omit it from the notation when no confusion arises.
     Since the map $x\rightarrow x^2$ is a bijection of $\F_{q^2},$ it suffices to consider $W_{f_\lambda}(\gamma^2)$ as $\gamma$ runs through $\F_{2^n}.$  For $\gamma=\gamma^2=0,$ the
balancedness of $f_\lambda$ gives $W_{f_\lambda}(0)=0$.

Suppose that $\gamma\ne0$.  Note that $P_\lambda(x)=x^2h(x^{2^m-1}),$ where $h(x)=x^3+\lambda x^2+\lambda.$ Then by Lemma \ref{lem1} and \ref{lemw1}, we have 
\begin{align}
\nonumber W_{f_\lambda}(\gamma^2)&=q-\sum_{z\in U_m}\sum_{y\in \F_{q}}\chi_m(\Tr_m^n(\gamma^2z^2h(\bar{z}^2))y^2+y(z+\bar{z}))\\
\nonumber&=q-\sum_{z\in U_m}\sum_{y\in \F_{q}}\chi_m(\Tr_m^n(\gamma z(h(\bar{z}^2))^{1/2})+(z+\bar{z}))y)\\
\label{eq12}&=q(1-\#\{z\in U_m: \psi_{\gamma}(z)=0\}),
\end{align}
where
\begin{equation}\label{eq27}
    \psi_{\gamma}(z)
=z^4+
\frac{1+\lambda^{1/2}(\gamma+\bar\gamma)}{\bar\gamma}(z^3+z)
+\frac{\gamma}{\bar\gamma}
\end{equation}

 Hence
$$W_{f_\lambda}(\gamma^2)\in \{q, 0, -q, -2q, -3q\}.$$

Note that $W_{f_\lambda}(0)=0$ and (\ref{eq12}) holds when $\gamma\ne 0,$ then for $i\ne 1,$ by (\ref{eq12}), we have
\begin{equation}\label{eq25}
    w_{q(1-i)}=\#\{\gamma\in \F_{q^2}^*:\psi_{\gamma}(z) \text{ has exactly } i \text{ distinct roots in } U_m\}.
\end{equation}

Since $f_\lambda(0)=1,$ by (\ref{eq2.1}), we have 
\[
\begin{cases}
w_q+w_0+w_{-q}+w_{-2q}+w_{-3q}=q^2,\\
qw_q-qw_{-q}-2qw_{-2q}-3qw_{-3q}=-q^2,\\
q^2w_q+q^2w_{-q}+4q^2w_{-2q}+9q^2w_{-3q}=q^4.
\end{cases}
\]
Solving for $w_q$, $w_0$, and $w_{-3q}$ gives
\begin{equation}\label{eq30}
    \begin{cases}
\displaystyle
w_q=\frac{q^2-3q}{4}
       +\frac{w_{-q}+w_{-2q}}{2},\\[2mm]
\displaystyle
w_0=\frac{2q^2+2q-4w_{-q}-3w_{-2q}}{3},\\[2mm]
\displaystyle
w_{-3q}
=\frac{q^2+q-2w_{-q}-6w_{-2q}}{12}.
\end{cases}
\end{equation}
It remains to determine $w_{-q}$ and $w_{-2q}$.

We now determine $w_{-2q}$. By (\ref{eq25}), this requires counting the number of $\gamma\in\mathbb \F_{q^2}^*$ for which $\psi_{\gamma}(z)$ has exactly three distinct roots in $U_m$. We claim that this occurs precisely when $\gamma\in T$, where $T=\{u:\operatorname{Tr}_1^m(u)=1\}$, which yields $w_{-2q}=q/2$.

Indeed, by Lemma~\ref{lem3.2.4}, $\psi_{\gamma}(z)$ has exactly three distinct roots in $U_m$ if and only if there exists $s\in U_m\setminus\{1\}$ such that
\begin{equation}\label{eq26}
    \psi_{\gamma}(z)
=(z+1)^2(z+s)(z+s^{-1}).
\end{equation}
If (\ref{eq26}) holds for some $s\in U_m\setminus\{1\}$, comparing coefficients gives $\gamma^{-1}=s+s^{-1},$ and hence
$\gamma=({s+s^{-1}})^{-1}\in T.$

Conversely, if $\gamma\in T$, then $\gamma\in\mathbb F_q$ and $\gamma^{-1}=s+s^{-1}$ for some $s\in U_m\setminus\{1\}$. Hence
\[
\psi_{\gamma}(z)=z^4+(s+s^{-1})(z^3+z)+1=(z^2+1)(z+s)(z+s^{-1}).
\]
This proves that $w_{-2q}=q/2$, as claimed.

We next determine $w_{-q}$. By (\ref{eq25}), this requires counting the number of $\gamma\in\mathbb \F_{q^2}^*$ for which $\psi_{\gamma}(z)$ has exactly two distinct roots in $U_m$. If $\gamma\in\F_q^*$, then
\[
\psi_{\gamma}(z)
=(z+1)^2(z^2+\gamma^{-1}z+1),
\]
which has either one or three distinct roots in $U_m$. Hence 
$\gamma\notin\F_q$ whenever $\psi_{\gamma}$ has exactly two distinct roots in $U_m$.

 In the following we assume that $\gamma\in\F_{q^2}\setminus\F_q$ and let $K_{\gamma_2}=1+\lambda^{1/2}(\gamma+\bar{\gamma}).$ By (\ref{eq27}), the equation
$\psi_{\gamma}(z)=0$ is equivalent to
\begin{equation}\label{eq28}
    \bar\gamma z^4+K_{\gamma_2}(z^3+z)+\gamma=0.
\end{equation}

Since $\psi_\gamma(1)=\frac{\bar{\gamma}+\gamma}{\bar\gamma}\ne0$, we only need to consider the roots of $\psi_\gamma(z)$ in $U_m\setminus\{1\}$. Choose $a$ and $\theta$ as in Lemma~\ref{lem3.2.3}, then for every $z\in U_m\setminus\{1\}$, Lemma~\ref{lem3.2.3} shows that there exists a unique $x\in\F_q$ such that
\[
z=\varphi(x)
=\frac{x+\theta}{\sqrt{D(x)}},
\]
where  $D(x)=x^2+x+a+1.$ 
Substituting $z=\varphi(x)$ directly into (\ref{eq28}), we obtain 
\begin{align*}
0
&=
\bar\gamma
 \left(\frac{x+\theta}{\sqrt{D(x)}}\right)^4
+K_{\gamma_2}
 \left[
 \left(\frac{x+\theta}{\sqrt{D(x)}}\right)^3
 +\frac{x+\theta}{\sqrt{D(x)}}
 \right]
+\gamma\\
&=
\frac{
 \bar\gamma(x+\theta)^4
 +K_{\gamma_2}
  \bigl((x+\theta)^3+(x+\theta)D(x)\bigr)\sqrt{D(x)}
 +\gamma D(x)^2
}{
 D(x)^2
}.
\end{align*}
Simplifying the numerator, the preceding equation reduces to
\[
\bar\gamma(x+\theta)^4
+K_{\gamma_2}(x+\theta)^2\sqrt{D(x)}
+\gamma D(x)^2=0.
\]
Moreover, recall that $D(x)=(x+\theta)(x+\theta+1).$
Hence
\begin{align*}
0
&=
\bar\gamma(x+\theta)^4
+K_{\gamma_2}(x+\theta)^2\sqrt{D(x)}
+\gamma D(x)^2\\
&=
(x+\theta)^2
\left(
 \bar\gamma(x+\theta)^2
 +K_{\gamma_2}\sqrt{D(x)}
 +\gamma(x+\theta+1)^2
\right).
\end{align*}
Since $x\in\F_q$ and $\theta\notin\F_q$, we have
$x+\theta\ne0$. Therefore,
\begin{equation}
\bar\gamma(x+\theta)^2
+K_{\gamma_2}\sqrt{D(x)}
+\gamma(x+\theta+1)^2=0.
\label{eq:unit-circle-reduction}
\end{equation}
Note that for each $\gamma\in \F_{q^2}\setminus\F_q,$ there exists a unique pair $(\gamma_1,\gamma_2)\in \F_q\times \F_q^*$ such that $\gamma=\gamma_1+\gamma_2\theta.$ Since $\bar\theta=\theta+1$, we have $\bar\gamma=\gamma_1+\gamma_2(\theta+1)$ and $\gamma+\bar\gamma=\gamma_2.$ Together with $\theta^2+\theta=a+1$, a direct calculation gives
\begin{align*}
\bar\gamma(x+\theta)^2+\gamma(x+\theta+1)^2=\gamma_2x^2+\gamma_2(a+1)+\gamma_1.
\end{align*}
Consequently, \eqref{eq:unit-circle-reduction} is equivalent to
\begin{align}
&\nonumber\gamma_2x^2+\gamma_2(a+1)+\gamma_1
+K_{\gamma_2}\sqrt{D(x)}=0,\\
\nonumber\Longleftrightarrow&\gamma_2^2x^4+\gamma_2^2(a+1)^2+\gamma_1^2+K_{\gamma_2}^2D(x)=0,\\
\label{eq29}\Longleftrightarrow&x^4
+\frac{K_{\gamma_2}^2}{\gamma_2^2}(x^2+x)
+\frac{\Delta(\gamma_1,\gamma_2)}
       {\gamma_2^2}=0,
\end{align}
where 
\[
\Delta(\gamma_1,\gamma_2)
=
\gamma_1^2+\gamma_2^2(a+1)^2
+K_{\gamma_2}^2(a+1).
\]
For ease of reference, denote the polynomial on the left-hand side of \eqref{eq29} by $H_{\gamma_1,\gamma_2}(x),$ and 
$$  A_{\gamma_2}
=\frac{K_{\gamma_2}^2}{\gamma_2^2},
\qquad
C_{\gamma_1,\gamma_2}
=\frac{\Delta(\gamma_1,\gamma_2)}{\gamma_2^2}.$$
Then
\[
H_{\gamma_1,\gamma_2}(x)
=x^4+A_{\gamma_2}(x^2+x)
 +C_{\gamma_1,\gamma_2}.
\]
If $K_{\gamma_2}=0$, then $A_{\gamma_2}=0$, and hence
\[
H_{\gamma_1,\gamma_2}(x)
=x^4+C_{\gamma_1,\gamma_2}.
\]
Since $x\mapsto x^4$ is a permutation of $\F_q$, this polynomial
has exactly one distinct root in $\F_q$. Therefore, such a value
of $\gamma_2$ does not contribute to $w_{-q}$.

Assume henceforth that $K_{\gamma_2}\ne0$, so that
$A_{\gamma_2}\ne0$. Let $g_{\gamma_2}(y)
=y^3+A_{\gamma_2}y+A_{\gamma_2}.$ We begin by counting the elements $\gamma_2\in\mathbb F_q^*$ such that $g_{\gamma_2}(y)$ has a unique solution.

By Lemma~\ref{lem3.2.1}, $g_{\gamma_2}(y)$ has a unique root in $\F_q$ if
and only if $\Tr_m(A_{\gamma_2}+1)=1.$

Since
\begin{align*}
A_{\gamma_2}+1=\frac{(1+\lambda^{1/2}\gamma_2)^2}{\gamma_2^2}+1=\gamma_2^{-2}+\lambda+1,
\end{align*}
we have
\begin{align}
\Tr_m(A_{\gamma_2}+1)
=
\Tr_m(\gamma_2^{-1})
+\Tr_m(\lambda)+\Tr_m(1).
\label{eq:resolvent-trace}
\end{align}

Suppose first that $\lambda\ne0$. Since
$\lambda\in\Lambda_m$, the polynomial $x^3+x+\lambda^{-1}$ has no root in $\F_q$. By Lemma~\ref{lem3.2.1}, we obtain
$\Tr_m(\lambda^2+1)=0.$ It follows that $\Tr_m(\lambda)=\Tr_m(1).$

Therefore, \eqref{eq:resolvent-trace} reduces to
\[
\Tr_m(A_{\gamma_2}+1)
=\Tr_m(\gamma_2^{-1}).
\]
Thus $g_{\gamma_2}$ has a unique root precisely when
$\Tr_m(\gamma_2^{-1})=1,$
subject to the additional condition $K_{\gamma_2}\ne0$.

There are $q/2$ elements $\gamma_2\in\F_q^*$ satisfying
$\Tr_m(\gamma_2^{-1})=1$. The equation
$K_{\gamma_2}=0$ has the unique solution
$\gamma_2=\lambda^{-1/2}.$ Note that 
$$\Tr_m(\lambda^{1/2})
=\Tr_m(\lambda)
=\Tr_m(1).$$
Thus, when $m$ is even, there are $q/2$ elements $\gamma_2\in\mathbb F_q^*$ for which $g_{\gamma_2}(y)$ has a unique root; when $m$ is odd, this number is $q/2-1$.

If $\lambda=0$, then $\lambda^{1/2}=0$ and
$K_{\gamma_2}=1$ for every $\gamma_2\in\F_q^*$. In this case,
\eqref{eq:resolvent-trace} becomes
\[
\Tr_m(\gamma_2^{-1})+\Tr_m(1)=1.
\]
When $m$ is odd, this is equivalent to
$\Tr_m(\gamma_2^{-1})=0$, which has $q/2-1$ solutions in
$\F_q^*$. When $m$ is even, it is equivalent to
$\Tr_m(\gamma_2^{-1})=1$, which has $q/2$ solutions.

To summarize,  we define
\[
\Gamma
:=
\left\{
\gamma_2\in\F_q^*:
K_{\gamma_2}\ne0,\
\Tr_m(A_{\gamma_2}+1)=1
\right\}.
\]
The preceding discussion gives
 $\Gamma=\{\gamma_2\in\F_q^*:
K_{\gamma_2}\ne0 \text{ and } g_{\gamma_2}(y) \text{ has a unique root in }\F_q\}$ and 
\begin{equation}
|\Gamma|
=
\begin{cases}
q/2-1,&m\ \mathrm{odd},\\
q/2,&m\ \mathrm{even}
\end{cases}
\label{eq:Gamma-lambda}
\end{equation}

Fix $\gamma_2\in\Gamma$, and denote by
$r(\gamma_2)$ the unique root of $g_{\gamma_2}(y)$ in $\F_q$. We now count the number of $\gamma_1\in\mathbb F_q$ for which $H_{\gamma_1,\gamma_2}(x)$ has exactly two distinct roots in $\mathbb F_q$.

Suppose first that $C_{\gamma_1,\gamma_2}=0$. This condition is equivalent to
\[
\gamma_1^2=\gamma_2^2(a+1)^2+K_{\gamma_2}^2(a+1),
\]
which has a unique solution $\gamma_1\in\mathbb F_q$. For this unique value, we have
\[
H_{\gamma_1,\gamma_2}(x)=x^4+A_{\gamma_2}(x^2+x)=x\bigl(x^3+A_{\gamma_2}x+A_{\gamma_2}\bigr)=xg_{\gamma_2}(x).
\]
Since $\gamma_2\in\Gamma$, $g_{\gamma_2}$ has the unique root $r(\gamma_2)$ in $\mathbb F_q$, and $A_{\gamma_2}=g_{\gamma_2}(0)\ne0$, so $r(\gamma_2)\ne0$. Hence $H_{\gamma_1,\gamma_2}$ has exactly the two distinct roots $0$ and $r(\gamma_2)$. Thus this case contributes exactly one element $\gamma_1$.

Now assume $C_{\gamma_1,\gamma_2}\ne0$. By Lemma~\ref{lem3.2.2} and $\gamma_2\in \Gamma$, $H_{\gamma_1,\gamma_2}(x)$ has exactly two distinct roots in $\mathbb F_q$ if and only if
\begin{equation}
\Tr_m\!\left(\frac{C_{\gamma_1,\gamma_2}\,r(\gamma_2)^2}{A_{\gamma_2}^2}\right)=0.
\label{eq:quartic-trace}
\end{equation}
Since $\gamma_2$ is fixed, the left-hand side of \eqref{eq:quartic-trace} is a function of $\gamma_1$. A direct expansion gives
\[
\Tr_m\!\left(\frac{C_{\gamma_1,\gamma_2}r(\gamma_2)^2}{A_{\gamma_2}^2}\right)
=
\Tr_m\!\left(\frac{r(\gamma_2)\gamma_2}{K_{\gamma_2}^2}\,\gamma_1\right)+\kappa_{\gamma_2},
\]
where $\kappa_{\gamma_2}\in\mathbb F_2$ is independent of $\gamma_1$. 
Since $A_{\gamma_2}\ne0$ and $r(\gamma_2)\ne0$, the coefficient of $\gamma_1$ is nonzero; hence, among all $\gamma_1\in\mathbb F_q$, exactly $q/2$ elements satisfy \eqref{eq:quartic-trace}.

We must exclude the one element $\gamma_1$ for which $C_{\gamma_1,\gamma_2}=0$, since it does not belong to the current case. Therefore, in the case $C_{\gamma_1,\gamma_2}\ne0$, the number of $\gamma_1$ for which $H_{\gamma_1,\gamma_2}$ has exactly two distinct roots is $q/2-1$.

Adding the unique solution from the case $C_{\gamma_1,\gamma_2}=0$, we conclude that for each fixed $\gamma_2\in\Gamma$, the total number of $\gamma_1\in\mathbb F_q$ such that $H_{\gamma_1,\gamma_2}(x)$ has exactly two distinct roots is $(q/2-1)+1=q/2.$ Hence $w_{-q}=\frac{q}{2}\cdot\#\Gamma.$ More precisely, we have 
\[
w_{-q}
=
\begin{cases}
\displaystyle\frac{q(q-2)}4,&m\ \mathrm{odd},\\[2mm]
\displaystyle\frac{q^2}4,&m\ \mathrm{even}.
\end{cases}
\]
Upon substituting the values of $w_{-q}$ and $w_{-2q}$ into \eqref{eq30}, a direct computation gives the Walsh spectrum of $f_\lambda$ as listed in Theorem~\ref{thm:lambdaWalsh}.
\hfill\ensuremath{\mbox{$\Box$}}
\end{proof}

Consequently, we obtain the nonlinearity of $f_\lambda.$
\begin{corollary}
Let $m\ge2$, $q=2^m$, and $\lambda\in\Lambda_m$. Then
\[
\mathcal{N}\mathcal{L}(f_\lambda)
=
\begin{cases}
4,&m=2,\\[1mm]
\displaystyle\frac{q^2}{2}-\frac{3q}{2},&m\ge3.
\end{cases}
\]
\end{corollary}

For $\lambda=1$, the polynomial $P_\lambda$ becomes $P_1(x)=x^2+x^{2q}+x^{3q-1}$, whose permutation property was studied by Gupta and Sharma \cite{gupta2016some}. Theorem~\ref{thm:lambdaWalsh} then specializes to the following.

\begin{theorem}\label{thm2}
Let $m\ge2$ with $\gcd(m,3)=1$, $q=2^m$,  and
let $f_1$ be the balanced Boolean function parameterized by $P_1(x+x^2)$.
Then the complete Walsh distribution of $f_1$ is given as follows.

If $m$ is odd, the spectrum is
\[
\{q,\ 0,\ -q,\ -2q,\ -3q\},
\]
where
\[
w_q=3\cdot\frac{q}{4}\left(\frac{q}{2}-1\right),\quad
w_0=\frac{q(2q+5)}{6},\quad
w_{-q}=\frac{q}{2}\left(\frac{q}{2}-1\right),\quad
w_{-2q}=\frac{q}{2},\quad
w_{-3q}=\frac{q(q-2)}{24}.
\]

If $m$ is even, the spectrum is
\[
\{q,\ 0,\ -q,\ -2q,\ -3q\},
\]
with
\[
w_q=\frac{q(3q-4)}{8},\quad
w_0=\frac{q(2q+1)}{6},\quad
w_{-q}=\frac{q^2}{4},\quad
w_{-2q}=\frac{q}{2},\quad
w_{-3q}=\frac{q(q-4)}{24}.
\]
\end{theorem}
\begin{example}
Similarly, Table \ref{tab:five-valued} presents several explicit examples of the Boolean functions $f_1$ defined in Theorem \ref{thm2}. The table gives their trace representations, Walsh spectral distributions, nonlinearity, algebraic degree, and algebraic immunity.

As in the four-valued case, although the univariate trace representation of $f_1$ becomes more complicated as $n$ increases, its Walsh spectrum is still completely determined by Theorem \ref{thm2}. 
Moreover, the numerical results indicate that these functions have relatively high algebraic degree and algebraic immunity close to the optimal upper bound $\lceil n/2\rceil$.

    \begin{table}[htbp]
\centering
\caption{Five-valued Walsh spectrum Boolean functions defined in Theorem \ref{thm2}}\label{tab:five-valued} 
\makebox[\textwidth][c]{
\begin{threeparttable}
\begin{tabular}{cccccc}
\toprule
$n$ & $f_1(x)=\operatorname{Tr}_n(R_n(x))+1$ & Walsh spectrum & $\mathcal{N}\mathcal{L}(f_1)$ & $\deg(f_1)$ & $\mathrm{AI}(f_1)$ \\
\midrule
 4 & $x^7+x$ & $\{0^6, 4^4 , -4^4, -8^2\}$ &  4& 3 & 2 \\
 8 & $x^{53}+x^{23}+x$ & $\{0^{88}, -32^{8}, 16^{88}, -16^{64}, -48^{8}\}$ &  104&  4 & 3 \\
 10 & $R_{10}(x)$\tnote{1} & $\{0^{368}, 32^{360},-96^{40},-64^{16}, -32^{240}\}$ & 464  &  6 & 4\\
 14 & $R_{14}(x)$\tnote{2} & $\{0^{5568}, 128^{6048}, -384^{672}, -256^{64}, -128^{4032}\}$ &8000  &  7 & 5 \\
\bottomrule
\end{tabular}
\begin{tablenotes}[flushleft]
\footnotesize
\item[1] $R_{10}(x)=x^{343}+x^{221}+x^{219}+x^{187}+x^{159}+x^{125}+x^{95}+x^{63}+x^{35}+x$

\item[2] $R_{14}(x)=x^{2731}+x^{2675}+x^{2669}+x^{2477}+x^{2421}+x^{1715}+x^{1653}+x^{1461}+x^{1401}+x^{1207}+x^{1147}+x^{953}+x^{699}+x^{637}
+x^{445}+x^{389}+x^{191}+x^{135}+x$

\end{tablenotes}
\end{threeparttable}}
\end{table}
\end{example}

\section{Applications to coding theory and finite geometry}\label{sec5}
The complete five-valued Walsh distribution has direct consequences
for few-weight codes, finite projective geometry, and algebraic graph
theory. Throughout this section, let $m\geq3$, $q=2^m$,
$\lambda\in\Lambda_m$, and let $f=f_\lambda$ be the Boolean function
defined in Theorem~3.10. Put
\[
D=\operatorname{Supp}(f),\qquad D^*=D\setminus\{0\}.
\]
Since $f$ is balanced and $f(0)=1$, we have
\[
|D|=\frac{q^2}{2},\qquad |D^*|=\frac{q^2}{2}-1.
\]

\subsection{A projective minimal five-weight code}
Define the binary linear code
\begin{equation}
\mathcal C_D=
\left\{
\bigl(\Tr_{1}^{2m}(bd)\bigr)_{d\in D^*}:b\in\F_{q^2}
\right\}.
\label{eq:CD}
\end{equation}

\begin{theorem}\label{thm:fiveweightcode}
The code $\mathcal C_D$ is a binary
\[
\left[\frac{q^2}{2}-1,\ 2m,\ \frac{q(q-3)}4\right]
\]
five-weight code. Its five nonzero weights are
\[
\frac{q(q-3)}4,\quad
\frac{q(q-2)}4,\quad
\frac{q(q-1)}4,\quad
\frac{q^2}4,\quad
\frac{q(q+1)}4.
\]
If $m$ is odd, the corresponding multiplicities are
\[
\frac{q(q-2)}{24},\quad
\frac q2,\quad
\frac{q(q-2)}4,\quad
\frac{q(2q+5)}6-1,\quad
\frac{3q(q-2)}8.
\]
If $m$ is even, they are
\[
\frac{q(q-4)}{24},\quad
\frac q2,\quad
\frac{q^2}4,\quad
\frac{q(2q+1)}6-1,\quad
\frac{q(3q-4)}8.
\]
\end{theorem}

\begin{proof}
For any nonzero $b\in\mathbb F_{q^2}$, define
\[
N_0:=\#\{d\in D:\operatorname{Tr}_1^{2m}(bd)=0\},\qquad
N_1:=\#\{d\in D:\operatorname{Tr}_1^{2m}(bd)=1\}.
\]
By the definition of $W_f(b)$, we have
\[
W_f(b)=\sum_{x\in\mathbb F_{q^2}}(-1)^{\operatorname{Tr}_1^{2m}(f(x)+bx)}
=-2\sum_{d\in D}(-1)^{\operatorname{Tr}_1^{2m}(bd)}
=-2(N_0-N_1).
\]
Since $N_0+N_1=|D|$, it follows that
\[
N_1=\frac{|D|}{2}+\frac14 W_f(b)=\frac{q^2}{4}+\frac14 W_f(b),
\]
Note that $\wt(c_b)=\#\{d\in D^*: \Tr_1^{2m}(bd)=1\}.$ Since $0\in D$ and $\Tr_1^{2m}(b\cdot 0)=0,$ it follows that 
\begin{equation}\label{eq5.1}
    \wt(c_b)=N_1=\frac{q^2}{4}+\frac{1}{4}W_f(b).
\end{equation} 

Substituting the five Walsh values from Theorem~\ref{thm:lambdaWalsh} into
\eqref{eq5.1} yields the five displayed nonzero weights.
Since $q\geq8$, all these weights are positive. Hence $c_b\ne0$
for every $b\in\F_{q^2}^*$, and therefore the linear map $b\longmapsto c_b$ is injective. Since it is surjective onto $\mathcal C_D$ by
definition, we obtain
\[
\dim_{\F_2}(\mathcal C_D)
=
\dim_{\F_2}(\F_{q^2})
=
2m.
\]

The multiplicities of the nonzero weights now follow from the Walsh
distribution in Theorem~3.10. The only exception is the weight
$q^2/4$, corresponding to the Walsh value $0$: the zero-Walsh class
contains $b=0$, which produces the zero codeword. Hence the
multiplicity of $q^2/4$ is $w_0-1$.

\hfill\ensuremath{\mbox{$\Box$}}
\end{proof}

\begin{corollary}\label{cor:minimalcode}
The code $\mathcal C_D$ is projective and minimal.
\end{corollary}

\begin{proof}
Let $\{\alpha_1,\ldots,\alpha_{2m}\}$ be an $\F_2$-basis of
$\F_{q^2}$, and let $\{\beta_1,\ldots,\beta_{2m}\}$ be its dual
basis with respect to the trace pairing, so that $\Tr_1^{2m}(\beta_i\alpha_j)=\delta_{ij}.$
Taking the codewords $c_{\beta_1},\ldots,c_{\beta_{2m}}$ as the
rows of a generator matrix, the column indexed by $d\in D^*$ is
\[
\bigl(\Tr_1^{2m}(\beta_1d),\ldots,
      \Tr_1^{2m}(\beta_{2m}d)\bigr)^{\mathsf T},
\]
which is precisely the coordinate vector of $d$ with respect to
the basis $\{\alpha_1,\ldots,\alpha_{2m}\}$.

Hence all columns are nonzero and pairwise distinct. Since the
base field is $\F_2$, two nonzero columns are proportional if and
only if they are equal. Therefore, $\mathcal C_D$ is projective.

Moreover,
\[
\frac{w_{\min}}{w_{\max}}
=
\frac{q(q-3)/4}{q(q+1)/4}
=
\frac{q-3}{q+1}
>
\frac12
\]
for $q\geq8$. By the Ashikhmin--Barg criterion~\cite{ashikhmin1998minimal},
every nonzero codeword of $\mathcal C_D$ is minimal. Therefore,
$\mathcal C_D$ is a projective minimal code.
\end{proof}

\subsection{Five hyperplane intersection numbers set}
Since the base field is binary, the nonzero vectors of $\F_{q^2}$ are naturally the points
of $\PG(2m-1,2)$. Thus $D^*$ may be regarded as a projective point set.
For $b\ne0$, let
\[
H_b=\{x\in\F_{q^2}:\Tr_1^{2m}(bx)=0\}
\]
be the corresponding projective hyperplane.

\begin{theorem}\label{thm:fivecharacter}
The set $D^*\subset\PG(2m-1,2)$ has exactly five hyperplane intersection numbers:
\[
\frac{q^2-q-4}{4},\quad
\frac{q^2-4}{4},\quad
\frac{q^2+q-4}{4},\quad
\frac{q^2+2q-4}{4},\quad
\frac{q^2+3q-4}{4}.
\]
The number of hyperplanes attaining each intersection number equals the multiplicity of the corresponding Walsh value $q,0,-q,-2q,-3q$, respectively, with the element $b=0$ removed from the zero-Walsh class. In particular, $D^*$ is a five-character point set with respect to hyperplanes.
\end{theorem}
\begin{proof}
For each nonzero $b$, the definitions of $H_b$ and $D^*$ give
\[
D^*\cap H_b=\{d\in D^*:(c_b)_d=0\},
\]
where $c_b$ denotes the corresponding codeword in $\mathcal C_D$ from Theorem~\ref{thm:fiveweightcode}. Consequently,
\[
|D^*\cap H_b|=|D^*|-\wt(c_b).
\]
Theorem~\ref{thm:fiveweightcode} then yields the five displayed intersection numbers and their respective frequencies.
\hfill\ensuremath{\mbox{$\Box$}}
\end{proof}

\subsection{A six-eigenvalue cubelike Cayley graph}
Let
\[
\Gamma_D=\Cay((\F_{q^2},+),D^*).
\]
Since the characteristic is two, $-D^*=D^*$, so $\Gamma_D$ is a simple undirected
regular Cayley graph.

\begin{corollary}\label{cor:cayley}
The graph $\Gamma_D$ has degree $q^2/2-1$ and exactly six distinct eigenvalues:
\[
\frac{q^2}{2}-1,
\quad -\frac q2-1,
\quad -1,
\quad \frac q2-1,
\quad q-1,
\quad \frac{3q}{2}-1.
\]
\end{corollary}

\begin{proof}
For the additive character $\chi_b(x)=(-1)^{\Tr_1^{2m}(bx)}$, the corresponding
eigenvalue is
$\sum_{d\in D^*}\chi_b(d).$
For $b=0$ this is $|D^*|$. For $b\ne0$,
\[
\sum_{d\in D^*}\chi_b(d)=-1-\frac12W_f(b).
\]
Substituting the five Walsh values gives the result. \hfill\ensuremath{\mbox{$\Box$}}
\end{proof}

\section{Balanced Boolean Functions with Simple Trace Representations from $P(x+x^{-1})$}\label{sec4}
In the preceding two sections, we discussed the construction of balanced Boolean functions parameterized by $P(x+x^2)$, their complete Walsh spectra, and their related applications. In this section, we turn to the two-to-one mapping $x+x^{-1}$. More precisely, by using the permutation polynomials given in Proposition \ref{prop3}, we construct an infinite family of balanced Boolean functions parameterized by $P(x+x^{-1})$. A main feature of this construction is that the resulting Boolean functions have very simple explicit trace representations.

We begin with a basic lemma describing the image set of the mapping $x+x^{-1}$.

\begin{lemma}\label{lem4.1}
Let $\F_{2^n}$ be the finite field with $2^n$ elements, and $T:=\{x+x^{-1}:x\in \F_{2^n}\}.$ Then
\begin{equation}\label{eq23}
T=\{a\in\F_{2^n}:\Tr_n(a^{-1})=0\}.
\end{equation}

Consequently, for every $x\in\F_{2^n}$, we have
$$
\Tr_n\left((x+x^{-1})^{-1}\right)=0.
$$
\end{lemma}

\begin{proof}
   It is clear that $0$ belongs to both sides of (\ref{eq23}), since $0=0+0^{-1}$ and $\Tr_n(0^{-1})=0$.
   
   Now let $a\in\F_{2^n}^*$. Then $a\in T$  if and only if the equation $x+x^{-1}=a$ has a solution in $\F_{2^n}$. Since $a\neq 0$, this is equivalent to $x^2+ax+1=0$. Dividing by $a^2$ and setting $y=x/a$,  we obtain that $a\in T$ if and only if  $y^2+y=a^{-2}$ has a solution $y\in \F_{2^n}.$
   
   By the Artin--Schreier criterion, the equation $y^2+y=a^{-2}$ has a solution in $\F_{2^n}$ if and only if $\Tr_n(a^{-2})=\Tr_n(a^{-1})=0$.  Therefore, $a\in T$ if and only if $\Tr_n(a^{-1})=0$. Hence (\ref{eq23}) holds.
   \hfill\ensuremath{\mbox{$\Box$}}
\end{proof}

For convenience,  let $m$ be even and  $$ G_k(x)=x\left(1+x^{k(2^m-1)}+x^{-k(2^m-1)}\right)\in \F_{2^{2m}}[x]. $$ Let $g_k$ be the Boolean function parameterized by $G_k(x+x^{-1})$. By Proposition \ref{prop3}, the polynomial $G_k(x)$ is a permutation polynomial over $\F_{2^{2m}}$ for every even integer $m$. Hence $G_k(x+x^{-1})$ is a two-to-one mapping over $\F_{2^{2m}}$, and consequently $g_k$ is a balanced Boolean function. 
\begin{theorem}
Let $m$ be an even number and $n=2m.$ Then we have 
\begin{equation}\label{eq14}   
 g_1(x)=\Tr_{n}(x^{2^{m-1}-1})+1.
\end{equation}
Consequently, the algebraic degree of $g_1$ is $m-1$.
\end{theorem}
\begin{proof}
    Since  $m$ is even, we have $\gcd(2^{m-1}-1, 2^{n}-1)=2^{\gcd(m-1,n)}-1=1.$ Consequently, the map $x\mapsto x^{2^{m-1}-1}$ is a permutation over $\F_{2^{n}},$  implying that $\Tr_{n}(x^{2^{m-1}-1})$ is  balanced. Therefore, to verify (\ref{eq14}), it suffices to show
that 
\begin{equation}\label{eq13}
\Tr_{n}(a^{2^{m-1}-1})=0
\end{equation}
holds for any $a\in \im(G_1(x+x^{-1})).$  

First, $0\in \operatorname{Im}(G_1(x+x^{-1}))$, because
 $G_1(1+1^{-1})=G_1(0+0^{-1})=0$. Thus,
\eqref{eq13} clearly holds for $a=0$.

Next, We consider the nonzero elements of the set $\im(G_1(x+x^{-1})).$ Let $a\in \im(G_1(x+x^{-1}))\setminus \{0\}.$ By the definition of $G_1(x),$  there exists some   $x\in \F_{2^{n}}\setminus\F_{2}$ such that 
\begin{align*}
    a&=(x+x^{-1})+(x+x^{-1})^{2^m-1+1}+(x+x^{-1})^{-(2^m-1)+1}\\
\end{align*}
Let $u=x+x^{-1}, v=u^{2^m-1}$ and $w=1+v+v^{-1}.$ Then
$$a=u(1+v+v^{-1})=uw.$$ Then $a=uw$ and 
\begin{align*}
    \Tr_{n}(a^{2^{m-1}-1})=\Tr_n(a^{2^m-2})=\Tr_n\left(\frac{vw^{2^m-1}}{uw}\right).
\end{align*}
Since $v^{2^m + 1} = 1$, we have $v^{2^m} = v^{-1}$, which implies $w \in \mathbb{F}_{2^m}$.
 Then we obtain
$$\Tr_{n}(a^{2^{m-1}-1})=\Tr_n(\frac{v}{uw}).$$
We now rewrite the element $\frac{v}{uw}.$  Using $v = u^{2^m - 1}$ and $w = 1 + u^{2^m - 1} + u^{1 - 2^m}$,  we can compute  
$$\frac{v}{uw}=\frac{1}{u}+\frac{u^{-1}+u^{-2^m}}{w}=\frac{1}{u}+\frac{u^{-1}}{w}+\left(\frac{u^{-1}}{w}\right)^{2^m}.$$ Therefore, by Lemma \ref{lem4.1}, we get
$$\Tr_n(a^{2^{m-1}-1})
=
\Tr_n\left(\frac{1}{u}+\frac{u^{-1}}{w}+\left(\frac{u^{-1}}{w}\right)^{2^m}\right)
=0$$
as desired. \hfill\ensuremath{\mbox{$\Box$}}
\end{proof}

For general $k\geq 2,$ we establish the following result.
\begin{theorem}
    Let $n=2m$ with $m$ even and $k\geq 2$ be a positive integer. Then we have
    \begin{equation}\label{eq15}
        g_k(x)=\Tr_n(x^{2^{n-1}-1}+x^{k(2^m-1)-1}+x^{(k-1)(2^m-1)-1})+1.
    \end{equation}
\end{theorem}
\begin{proof}
    For convenience, we write $h_k(x)=\Tr_n(x^{2^{n-1}-1}+x^{k(2^m-1)-1}+x^{(k-1)(2^m-1)-1})+1.$ First, we show that $h_k(x)$ is balanced. It suffices to prove that $W_{h_k}(0)=0.$ Note 
    \begin{align*}
        W_{h_k}(0)&=-\sum_{x\in\F_{2^n}}\chi_n\left(x^{2^{n-1}-1}+x^{k(2^m-1)-1}+x^{(k-1)(2^{m}-1)-1}\right)\\
        &=-\sum_{x\in\F_{2^n}^*}\chi_n\left(x^{2^{n-1}-1}+x^{k(2^m-1)-1}+x^{(k-1)(2^{m}-1)-1}\right)-1\\
        &=-\sum_{z\in U_m}\sum_{y\in \F_{2^m}^*}\chi_n\left(y^{-1}(z^{-1}+z^{-2k-1}+z^{-2k+1})\right)-1\\
        &=-\sum_{z\in U_m}\sum_{y\in \F_{2^m}}\chi_m\left(y^{-1}(z^{-1}+z^{-2k-1}+z^{-2k+1}+z+z^{2k+1}+z^{2k-1})\right)+2^m\\
        &=-2^m\cdot\#\{z\in U_m : (z+z^{-1})(1+z^{2k}+z^{-2k})=0\}+2^m.
    \end{align*}

     Since every element of $U_m$ is nonzero,
the equation $z+z^{-1}=0$ is equivalent to
$z^2+1=0,$ and  it has the unique solution $z=1$ in $U_m$.
    
    On the other hand, note that $z^2+z+1=0$ has no solution in $U_m$. Indeed, as $m$ is even, we have $3\nmid 2^m+1$, so $U_m$ contains no primitive third roots of unity. Thus, if $1+z^{2k}+z^{-2k}=0$ for some $z\in U_m$, then $y=z^{2k}\in U_m$ would satisfy $y^2+y+1=0$, which is impossible. Therefore,
$$
W_{h_k}(0)=-2^m+2^m=0.
$$
Thus $h_k$ is balanced.

Since $g_k(x)$  is also balanced, in order to show (\ref{eq15}), it suffices to show that for any $a\in \im(G_k(x+x^{-1})),$  
\begin{equation}\label{eq16}
    h_k(a)+1=\Tr_n(a^{2^{n-1}-1}+a^{k(2^m-1)-1}+a^{(k-1)(2^m-1)-1})=0.
\end{equation}

For $a=0,$ (\ref{eq16}) holds clearly. Now we assume that $a\in\im(G_k(x+x^{-1}))\setminus\{0\}.$ Then there exists some $x\in \F_{2^n}$ such that 
$$a=(x+x^{-1})+(x+x^{-1})^{k(2^m-1)+1}+(x+x^{-1})^{-k(2^m-1)+1.}$$
Let $u=x+x^{-1}, v=u^{2^{m}-1}.$ We get 
$$a=u(1+v^k+v^{-k}).$$ Since $v=u^{(2^m-1)}, v^{2^m}=u^{2^n-2^m}=u^{1-2^m}=v^{-1}.$ It follows that $1+v^k+v^{-k}\in\F_{2^m}.$ Therefore we obtain
\begin{align*}
h_k(a)+1&=\Tr_n(a^{2^n-2}+a^{k(2^m-1)-1}+a^{{(k-1)(2^m-1)-1}})\\
&=\Tr_n\left(\frac{1}{u(1+v^k+v^{-k})}+\frac{v^k}{u(1+v^k+v^{-k})}+\frac{v^{k-1}}{u(1+v^k+v^{-k})}\right)\\
&=\Tr_n\left(\frac{1}{u}+\frac{v^{-k}+v^{k-1}}{u(1+v^k+v^{-k})}\right)\\
&=\Tr_n\left(\frac{v^{-k}+v^{k-1}}{u(1+v^k+v^{-k})}\right).
\end{align*}
The final equality follows from Lemma \ref{lem4.1}, which asserts that $\Tr_n\left(\frac{1}{u}\right) = \Tr_n\left(\frac{1}{x + x^{-1}}\right) = 0$.
Note that $(1+v^k+v^{-k})\in \F_{2^m}$ and
$$\left(\frac{v^{-k}}{u}\right)^{2^m}=\left(\frac{v^k}{u\cdot v}\right)=\frac{v^{k-1}}{u},$$ we obtain
\begin{align*}
h_k(a)+1&=\Tr_n\left(\frac{v^{-k}}{u(1+v^k+v^{-k})}+\frac{v^{k-1}}{u(1+v^k+v^{-k})}\right)\\
&=\Tr_n\left(\frac{v^{-k}}{u(1+v^k+v^{-k})}+\left(\frac{v^{-k}}{u(1+v^k+v^{-k})}\right)^{2^m}\right)\\
&=0.
\end{align*}
\hfill\ensuremath{\mbox{$\Box$}}
\end{proof}

\section{Conclusion}\label{sec6}

In this paper, we investigated balanced Boolean functions arising
from two-to-one mappings of the forms $P(x+x^2)$ and
$P(x+x^{-1})$, where $P$ is a sparse permutation polynomial over
$\F_{2^{2m}}$ with generalized Niho-type exponents.

For mappings of the form $P(x+x^2)$, we first determined the
complete four-valued Walsh distribution of the balanced Boolean
function associated with $Q(x)=x\left(1+x^{q-1}+x^{2(q-1)}\right),$
where $q=2^m$ and $m$ is even. The resulting function has Walsh values
$q,0,-q,-2q$ and nonlinearity $q^2/2-q$.

We also studied the parameterized family
$$
P_\lambda(x)=x^{3q-1}+\lambda x^{2q}+\lambda x^2,
$$
where $\lambda\in\F_q.$
We proved that $P_\lambda$ permutes $\F_{q^2}$ if and only if
either $\lambda=0$, or $\lambda\ne0$ and
$X^3+X+\lambda^{-1}$ has no root in $\F_q$.
Together with Bluher's counting result~\cite{bluher2004on},
this criterion gives $(q+2)/3$ admissible parameters when
$m$ is even and $(q+4)/3$ when $m$ is odd.
For every admissible $\lambda$, we determined the complete Walsh
distribution of the balanced Boolean function $f_\lambda$
parameterized by $P_\lambda(x^2+x)$. A notable feature of this
construction is that the complete Walsh distribution is independent
of the choice of $\lambda$. For $m\geq3$, all these functions have
the same five-valued Walsh spectrum $\{q,0,-q,-2q,-3q\},$
whereas for $m=2$, the value $-3q$ has multiplicity zero and the
spectrum degenerates to four values. 

The complete five-valued Walsh distribution also yields several
combinatorial applications. From the support of $f_\lambda$, we
obtained a family of projective minimal binary five-weight codes
with explicitly determined weight distributions. The same support
gives a five-character point set in $\operatorname{PG}(2m-1,2)$
with five explicitly determined hyperplane intersection numbers,
and it also defines a regular cubelike Cayley graph with exactly
six distinct eigenvalues. These consequences illustrate the close
interaction among few-valued Walsh spectra, few-weight codes,
finite projective geometry, and spectral graph theory.

For mappings of the form $P(x+x^{-1})$, we used the permutation
polynomials
\[
G_k(x)=x\left(1+x^{k(q-1)}+x^{-k(q-1)}\right)
\]
with even $m$ to construct an infinite family of balanced Boolean
functions. By directly characterizing the relevant image sets, we
obtained simple explicit trace representations for these functions,
providing a second application of the parameterization approach
beyond Walsh-spectrum evaluation.

Several questions remain open. It would be interesting to determine
whether the parameter-independent Walsh phenomenon exhibited by
$P_\lambda$ extends to broader classes of generalized Niho-type
permutation polynomials, and to identify structural conditions on
$P$ that force few-valued Walsh spectra. The coding and geometric
objects arising from the supports of $f_\lambda$ also merit further
study, including their automorphism groups, possible geometric
characterizations, and relations with other known families of
few-character sets and minimal codes. More generally, replacing
$x^2+x$ and $x+x^{-1}$ by other sparse two-to-one mappings may lead
to new classes of balanced Boolean functions whose spectral and
algebraic properties can still be determined explicitly.

\bibliographystyle{plain}	
\bibliography{References}
\end{document}